\documentclass[twoside]{article}

\usepackage{PRIMEarxiv}

\usepackage[utf8]{inputenc} % allow utf-8 input
\usepackage[T1]{fontenc}    % use 8-bit T1 fonts
\usepackage{hyperref}       % hyperlinks
\usepackage{url}            % simple URL typesetting
\usepackage{booktabs}       % professional-quality tables
\usepackage{amsfonts}       % blackboard math symbols
\usepackage{nicefrac}       % compact symbols for 1/2, etc.
\usepackage{microtype}      % microtypography
\usepackage{lipsum}
\usepackage{fancyhdr}       % header
\usepackage{graphicx}       % graphics
\graphicspath{{media/}}     % organize your images and other figures under media/ folder
\usepackage{amsmath} 
\usepackage{float}

\usepackage{enumitem}
\usepackage{soul}
\usepackage{multirow}
\usepackage{multicol}
\usepackage{xcolor}
\usepackage{array}

\newcommand{\rr}[1]{{\color{black}#1}} %blue

\definecolor{darkgreen}{rgb}{0,0.7,0}

\title{CC30k: A Citation Contexts Dataset for \\Reproducibility-Oriented Sentiment Analysis}

\author{
  Rochana R. Obadage \\
  Old Dominion University  \\
  Norfolk, VA, USA \\
  \texttt{rochana@cs.odu.edu} \\
  \And
  Sarah Rajtmajer  \\
  The Pennsylvania State University  \\
  University Park, PA, USA \\
  \texttt{smr48@psu.edu} \\
  \And
  Jian Wu \\
  Old Dominion University  \\
  Norfolk, VA, USA \\
  \texttt{j1wu@odu.edu} \\
}

\begin{document}
\maketitle

\begin{abstract}

Sentiments about the reproducibility of cited papers in downstream literature offer community perspectives and have shown as a promising signal of the actual reproducibility of published findings. To train effective models to effectively predict reproducibility-oriented sentiments and further systematically study their correlation with reproducibility, we introduce the CC30k dataset, comprising a total of 30,734 citation contexts in machine learning papers. Each citation context is labeled with one of three reproducibility-oriented sentiment labels: Positive, Negative, or Neutral, reflecting the cited paper’s perceived reproducibility or replicability. Of these, 25,829 are labeled through crowdsourcing, supplemented with negatives generated through a controlled pipeline to counter the scarcity of negative labels. Unlike traditional sentiment analysis datasets, CC30k focuses on reproducibility-oriented sentiments, addressing a research gap in resources for computational reproducibility studies. The dataset was created through a pipeline that includes robust data cleansing, careful crowd selection, and thorough validation. The resulting dataset achieves a labeling accuracy of 94\%. We then demonstrated that the performance of three large language models significantly improves on the reproducibility-oriented sentiment classification after fine-tuning using our dataset. The dataset lays the foundation for large-scale assessments of the reproducibility of machine learning papers. The CC30k dataset and the Jupyter notebooks used to produce and analyze the dataset are publicly available at \textit{\url{https://github.com/lamps-lab/CC30k}}.
\end{abstract}

% keywords can be removed

\section{Introduction}

\begin{figure}[ht]
\vspace{-0.1cm}
  \centering
  % [width=\linewidth, frame]{figure-filename}
    \setlength{\fboxsep}{2.5pt} % Adjust the padding
    \setlength{\fboxrule}{0.5pt} % Adjust the border width
  \fbox{\includegraphics[width=0.6
  \linewidth]{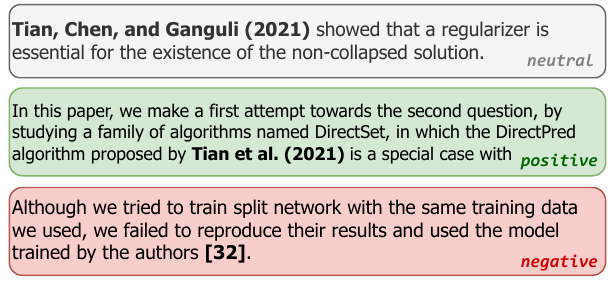}}
  \caption{Examples of citation context with different reproducibility-oriented sentiments \cite{acm-rep-24}.}%; examples from Tian et al. \cite{RS_062} and Chris et al. \cite{Chris-et-al}}
  \label{citaion_contexts_example}
\vspace{-0.3cm}
\end{figure}

Citation contexts are textual fragments in scholarly documents that surround and contain citations to prior work (see Fig. \ref{citaion_contexts_example}). Citation context can tell us why the work was cited, offer insight into the authors' perspectives on the work, and highlight relationships between the cited and citing works. Here, we explore one particular opportunity inherent to citation contexts, namely, the opportunity to mine citation contexts for signals of reproducibility. In computational fields such as Machine Learning (ML) and Artificial Intelligence (AI), researchers often share and reuse code to compare models against benchmark datasets or build upon and extend learning frameworks. In the process, they often need to reproduce (using the same methods and datasets) or replicate (using the same methods and different datasets) the results reported in another paper.\footnote{Throughout this work, we adopt the definition of \textbf{reproducibility} from \cite{national2019reproducibility} where a finding is deemed reproducible if \textit{consistent results are obtained using the same input data, computational steps, methods, and code, and conditions of analysis}. Similarly, we adopt the definition of \textbf{replicability} from \cite{national2019reproducibility}, which refers to \textit{obtaining consistent results across studies aimed at answering the same scientific question, with each study collecting its own data.} These definitions align with those adopted by ACM \cite{acmartifact}.} The outcome of this process can be reported in their paper as citation contexts. Despite norms of code and data sharing and researchers' advanced technical expertise, automatic reproducibility assessment of published work remains a pressing challenge for the computer sciences broadly \cite{Collberg2016, Hocquet2021-ou, Antunes-2024} and the AI community in particular \cite{Raff-NEURIPS2019, raff2024machinelearningresearchersmean, Gundersen-Kjensmo-2018, Gundersen2024}. It was empirically shown using a small dataset that these citation contexts may serve as a promising signal of the reproducibility of the cited study \cite{acm-rep-24}. Moreover, writ large, analyzing these contexts can help identify patterns, trends, and gaps in reproducibility to advance our understanding of scientific transparency and best practices \cite{Elliott-2022}.

To facilitate the study of reproducibility assessment and its relation with citation context, we develop the CC30k dataset comprising 30,734 citation contexts from scientific literature in ML (Table \ref{tab:rep_study_sources}). Each context identifies a cited paper through its citation mark and is labeled into one of three categories reflecting reproducibility-oriented sentiments (ROS): \textit{Positive}, \textit{Negative}, and \textit{Neutral}. % ----- %-------- %Methodology 
The CC30k dataset was created using a pipeline. We extracted citation contexts, parsed citation marks and sentence structures, and filtered citation contexts to eliminate ambiguity caused by multiple citation markers within the same context. We used crowdsourcing to annotate the contexts via \textit{Amazon Mechanical Turk} \cite{mturkAmazonMechanical}. Each context was labeled by three independent annotators, with majority voting employed to determine the final label. We implemented a comprehensive and careful worker selection mechanism, including using a pilot dataset and multiple metrics to build a pool of qualified annotators. Verifying the crowdsourced results indicates that it is possible to obtain high-quality ROS labels on citation contexts in AI papers without relying on annotators with domain-specific backgrounds.
Our contributions are as follows:
\begin{itemize}
    \item We introduce \textbf{CC30k}, a large-scale dataset of 30,734 citation contexts, each labeled according to its reproducibility-oriented sentiment. 
    \item We demonstrate why traditional sentiment classifiers fail to capture reproducibility-oriented sentiment. 
    \item To illustrate the utility of CC30k, we fine-tune several large language models (LLMs) on the dataset for the task of ROS classification. 
    \item We compare CC30k with existing datasets for both sentiment analysis and citation analysis. 
    \item We analyze and categorize citation contexts based on their citation mark formats. 
\end{itemize}

\section{Related Work}
Existing datasets in sentiment analysis and citation context analysis have yet to incorporate signals of reproducibility and replicability (see Table \ref{tab:dataset_comparison}). Traditional sentiment analysis datasets, such as IMDb Movie Reviews \cite{imdb-50k} or SemEval 2017 Task 4 \cite{semeval2017task4sentiment}, are designed for sentiment classification tasks in web-based text data such as product reviews \cite{customer-reviews} or social media posts \cite{social-media}. These datasets played a crucial role in advancing sentiment analysis models and enabling the development of general sentiment analysis tools. Many of these datasets have become core components in widely adopted software packages and are routinely featured on leaderboards such as those maintained on the \texttt{paperswithcode} \cite{paperswithcode} platform. 

Citation context datasets such as the ACL-ARC subset \cite{acl-arc-subset}, SciCite \cite{cohan:naacl19}, and the Critical Citation Contexts Corpus \cite{bordignon_2024_10694465} typically focus on broader citation functions (e.g., background, uses, compares or contrasts, motivation, continuation, future), citation intents (e.g., background, method, result), or on citations that indicate scholarly critique or debate, without specifically addressing reproducibility. While these resources have enriched bibliometric analyses and enabled a deeper understanding of citation roles, they lack the specificity required to evaluate reproducibility-oriented sentiments in scholarly texts. The CC30k dataset bridges these domains through identifying reproducibility-oriented sentiments in citation contexts.

\renewcommand{\arraystretch}{1.15}
\begin{table*}[h]
\small
\centering
\caption{Comparison of CC30k with existing datasets for sentiment (above the middle line) and citation analysis (below the middle line).}
\begin{tabular}{p{0.23\textwidth}>{\raggedleft\arraybackslash}p{0.05\textwidth}p{0.12\textwidth}p{0.26\textwidth}p{0.24\textwidth}}
\hline
\textbf{Dataset Name} & \textbf{Size} & \textbf{Data Category} & \textbf{Purpose} & \textbf{Applications} \\ \hline

Internet Movie Database (IMDb) \cite{imdb-50k} & 50,000 & reviews & Evaluate sentiment analysis models on complex movie reviews & Sentiment classification for movie reviews \\ %\hline
SemEval-2017 Task 4 \cite{semeval2017task4sentiment} & 65,000 & tweets & Comprehensive benchmark for various sentiment analysis subtasks & Sentiment classification, topic-specific sentiment, Tweet quantification \\ %\hline

Twitter US Airline Sentiment \cite{kaggleTwitterAirline} & 14,640 & tweets & Evaluate sentiment analysis models with imbalanced sentiment classes & Sentiment analysis for customer reviews of airlines \\ %\hline
Sentiment140 \cite{Sentiment140} & 1.6M & tweets & Evaluate models on large-scale real-world social media data & Sentiment analysis for short, informal social media posts \\ %\hline

SentiGrad \cite{SentiGrad} & 6,064 & youtube comments & Sentiment analysis of code-mixed comments & Educational content evaluation, code-mixed language processing \\ \hline

Critical Citation Contexts Corpus \cite{bordignon_2024_10694465}  & 505 & citation contexts & Study of critical citations to understand mechanisms and develop detection tools & Training tools for automatic retrieval of critical citations \\ %\hline
IMS Citation Corpus \cite{jochim-schutze-2012-towards} & 2,008 & citation contexts & Citation classification using faceted classification schemes & Bibliometrics and citation analysis \\ %\hline
Context-Enhanced Citation Sentiment Detection \cite{athar-teufel-2012-context} & 1,741 & citation contexts & Sentiment analysis of citation contexts considering broader contexts & Bibliometric search and citation sentiment detection \\ %\hline
3C Shared Task (2021) \cite{3c-shared-task-influence-v2} & 3,000 & citation contexts & Citation context classification based on influence & Classification of incidental vs. influential citations \\ %\hline

Citation FPAI \cite{HERNÁNDEZ-2017} & 2,120 & citation contexts & Citation analysis using a classification scheme based on function, polarity, aspects, and influence & Evaluation of citation roles and impact on citing papers \\ %\hline

SCiFi \cite{cao2024-scifi} & 10,000 & wikipedia citations & Verifiable generation with fine-grained citations & Precise citation generation, verifiable content creation \\

ACL-ARC Subset \cite{acl-arc-subset} & 1,969 & citation contexts & Citation intent classification (e.g., background, method, result) & Bibliometrics and citation intent studies \\ %\hline
SciCite \cite{cohan:naacl19} & 10,969 & citation contexts & Classify citation intents in academic papers & Research in citation intent classification \\ %\hline

MCG-S2ORC \cite{MCG-S2ORC} & 17,210 & citation contexts & Multi-sentence citation context generation & Citation context generation, academic writing assistance \\

\textbf{CC30k (ours)} & \textbf{30,734} & \textbf{citation contexts} & \textbf{Reproducibility-oriented sentiment analysis (Pos/Neg/Neutral)} & \textbf{Reproducibility prediction and assessment} \\ \hline

\end{tabular}
\label{tab:dataset_comparison}
\end{table*}

\section{Methods}
\subsection{Overview}
\begin{figure}[]
  \centering
  \setlength{\fboxsep}{0pt} % Adjust the padding
  \setlength{\fboxrule}{0.5pt} % Adjust the border width
  \fbox{\includegraphics[trim=14 12 48 12, clip, width=0.6\linewidth]{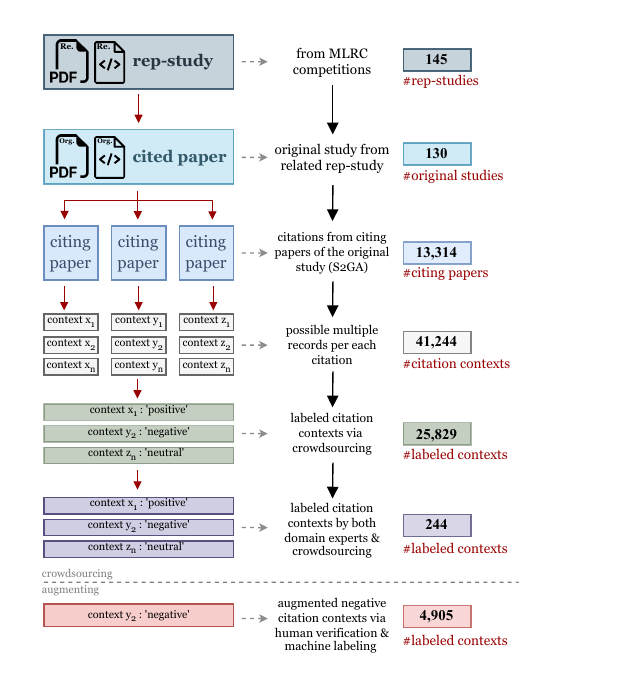}}
  \caption{The data production process.}
  \label{fig:citaion_contexts_relationships}
\end{figure}

The raw dataset contains 41,244 citation contexts without any ROS labels, introduced in our previous study \cite{acm-rep-24}. We built a small dataset consisting of 1,937 citation contexts whose ROS labels were manually assigned by domain experts. However, the dataset contained only 23 samples labeled with negative ROS, a tiny fraction of the labeled samples. Because manual labeling is time-consuming and thus does not scale well, we explored the assistance of large language models (LLMs). However, both LLMs and existing generic sentiment analysis models struggled to accurately label ROS (see Table \ref{tab:classification-reports}). Therefore, here we explore the crowdsourcing method. Fig. \ref{fig:citaion_contexts_relationships} provides a schematic illustration of the data preparation workflow.

\vspace{-10pt}
\subsection{Dataset Production}
The data production pipeline (Fig. \ref{fig:citaion_contexts_relationships}) contains several key terms: 
\begin{itemize}
    \item \textbf{Original studies (cited papers)}: Original research papers whose results are reproduced in reproducibility studies. 
    \item \textbf{Reproducibility studies (rep-studies)}: Papers documenting attempts to reproduce reported results from original studies.
    \item \textbf{Citing papers}: Papers citing original studies, by including a citation mark in a sentence, which is called \textbf{\textit{citation context}}. Here, we only study explicit citation context, which includes citation marks \cite{guo-etal-2020-smartcitecon}.
\end{itemize}
Data preparation included the following key steps:
\begin{enumerate}
    \item \textbf{Data collection}: We collected metadata and PDF files for reproducibility studies (145 rep-studies) and their associated original studies (130 original papers). We primarily sourced rep-studies from the Machine Learning Reproducibility Challenge (MLRC) competitions \cite{akella2023layingfoundationsquantifyeffort} and additional resources, as detailed in Table~\ref{tab:rep_study_sources}. 
    
    \item \textbf{Citation context collection}: We obtained citation contexts about original studies from citing papers using Semantic Scholar Graph API (S2GA) \cite{Kinney2023TheSS}, and identified a total of 13,314 citing papers referencing the original studies, resulting in 41,244 citation contexts (collected January 2024).
    \item \textbf{Crowdsourcing for ROS labeling}: After filtering out citation contexts with ambiguous citation marks (see sections \ref{lbl:citation-mark-analysis} and \ref{lbl:pre-processing-filtering}), we annotated citation contexts through crowdsourcing, yielding 25,829 labeled contexts, each categorized as \textit{Positive}, \textit{Negative}, or \textit{Neutral}.
    \item \textbf{Crowdsourced label verification}: To assess the quality of crowdsourced labels, we manually validated $\sim$1\% (244 citation contexts) of the dataset using stratified sampling.
    \item \textbf{Negative augmentation}: To address the class imbalance, we added 4,905 negative citation contexts by combining labels from human experts and a supervised machine learning labeler.

\end{enumerate}

\begin{table}[ht]
    \caption{Data sources for selected reproducibility studies with the year of reproduction, number of rep-studies (\#Rep-S.), number of original studies (\#Org-S.), number of citations (\#Cit.), and number of citation contexts (\#contexts).}
    \footnotesize
    \centering
    \begin{tabular}{lcccrr}
    \hline
    \textbf{\footnotesize Data Source}  & \textbf{\footnotesize Year} & \textbf{\footnotesize \#Rep-S.} & \textbf{\footnotesize \#Org-S.} & \textbf{\footnotesize \#Cit.} & \textbf{\footnotesize \#Contexts} \\
    \hline
        ICLR \cite{repromlReproducibilityChallenge,rescienceReScience} & 2019 & 4 & 4 & 651 & 2,102 \\
        %\hline
        NeurIPS \cite{repromlReproducibilityChallenge,rescienceReScience} & 2019 & 10 & 10 & 3,364 & 9,908 \\
        %\hline
        MLRC \cite{repromlReproducibilityChallenge,rescienceReScience} & 2020 & 23 & 22 & 3,224 & 10,798 \\
        %\hline
        MLRC \cite{repromlReproducibilityChallenge,rescienceReScience} & 2021 & 47 & 41 & 2,596 & 6,958 \\
        %\hline
        MLRC \cite{repromlReproducibilityChallenge,rescienceReScience} & 2022 & 45 & 37 & 2,881 & 9,869 \\
        %\hline
        TSR \cite{10.1007/978-3-031-41679-8_1} & 2023 & 16 & 16 & 598 & 1,609 \\
    \hline
        \textbf{Total} & & 145 & 130 & 13,314 & 41,244 \\
    \hline
    \end{tabular}
    \label{tab:rep_study_sources}
\end{table}

\subsection{Citation mark analysis}\label{lbl:citation-mark-analysis}

A \textbf{\textit{citation mark}}, or inline citation, refers to a complete entry in a bibliography elsewhere in the paper \cite{Inline-Citation}. It may appear in parentheses, brackets, or as a superscript number. Citation marks have many styles, usually associated with the bibliographic style adopted throughout the paper. For example, the citation marks associated with the APA style include up to three authors' last names followed by the publication year. In the IEEE style, it includes only an Arabic number in square brackets.

A citation context may cite several papers in several places. To avoid confusion for labelers in the crowdsourcing and ensure that only uniquely identifiable (e.g., Table \ref{tab:citation_mark_formats}: Items (a) and (f)) citation contexts were included in the samples to be annotated, we analyzed citation marks present within citation contexts and  exclude those with multiple, inconsistent, or ambiguous citation marks (e.g., Table \ref{tab:citation_mark_formats}: Items (c) and (e)) % would be unsuitable for labeling and analysis 
(citation context filtering processes are further discussed in Section \ref{lbl:pre-processing-filtering}). This process involves visually examining various citation formats, parsing citation marks using regular expressions, recognizing the appearance of citation marks in citation contexts, and establishing filtering rules. 

%\subsubsection{Diverse citation mark styles}
By programmatically examining the entire citation context collection (41,244), we found that the most common text-based formats in our dataset are MLA, APA, and IEEE. We categorize citation contexts based on the number of citation marks they contain:

\begin{itemize}[leftmargin=*]
    \item \textbf{Contexts with a single citation mark:} %These contexts contain a single citation reference, such as 
    For example, the IEEE format (e.g., \texttt{"[1]"}) or MLA/APA format (e.g., \textit{"(Han, 2022)"}).
    \item \textbf{Contexts with multiple citation marks:} %These contexts contain more than one citation mark. 
    Examples include multiple IEEE citations (e.g., \texttt{[1][2][3]}), or multiple MLA/APA citations (e.g., \textit{(Zhang et al., 2022; Höllein et al., 2022; Xie et al., 2022)}).
\end{itemize}

\renewcommand{\arraystretch}{1.25} % Increase row height by 50%
\begin{table*}[h]
\small
\centering
\caption{Identified citation mark formats via regular expression-based parsing and their descriptions. Contexts with formats (c) and (e) were excluded from this study.}
% \begin{tabular}{p{4cm}p{10cm}}
\begin{tabular}{p{0.04\textwidth} p{0.25\textwidth} p{0.64\textwidth}}
\hline
{\textbf{Item}}&\textbf{Citation Mark Format} & \textbf{Description} \\ \hline
{(a)}&\texttt{[1]} & A single numeric citation mark, commonly used in IEEE and similar styles. \\ %\hline
{(b)}&\texttt{[1][2][3]} & Multiple numeric citation marks appearing consecutively, representing multiple references. \\ %\hline
{(c)}&\texttt{[1] and [2]} & Numeric citation marks separated by ``and" indicating distinct references. \\ %\hline
{(d)}&\texttt{[1,2,3]} & A range of numeric citation marks listed together, representing multiple references within a single bracket. \\ %\hline
{(e)}&\texttt{[1,2,3]} text text text \texttt{[4]} & A combination of grouped numeric citation marks followed by another distinct numeric citation mark, often within the same context. \\ %\hline
{(f)}&(Yao, 2022) & A textual citation with a single author, often used in MLA or APA styles. \\ %\hline
{(g)}&(Yao et al., 2022) & A textual citation indicating multiple authors, abbreviated using \textit{et al.}. \\ %\hline
{(h)}&(Zhang et al., 2022; Höllein et al., 2022; Xie et al., 2022) & Multiple textual citations with \textit{et al.}, separated by semicolons. Often used in APA or similar styles. \\ %\hline
{(i)}&(Iwasawa \& Matsuo, 2021) & A textual citation listing two authors explicitly, separated by \texttt{\&}. Common in styles like APA. \\ %\hline
{(j)}&(Zhang et al., 2022; Iwasawa \& Matsuo, 2021; Xie et al., 2022) & A combination of textual citations, including both \textit{et al.} and explicit \texttt{\&}, separated by semicolons. \\ \hline
\end{tabular}
\label{tab:citation_mark_formats}
\end{table*}

%\subsubsection{Citation Contexts with One or Multiple Citation Marks}

% \begin{itemize}[leftmargin=*]
%     \item \textbf{Contexts with a single citation mark:} %These contexts contain a single citation reference, such as 
%     For example, the IEEE format (e.g., \texttt{"[1]"}) or MLA/APA format (e.g., \textit{"(Han, 2022)"}).
%     \item \textbf{Contexts with multiple citation marks:} %These contexts contain more than one citation mark. 
%     Examples include multiple IEEE citations (e.g., \texttt{[1][2][3]}), or multiple MLA/APA citations (e.g., \textit{(Zhang et al., 2022; Höllein et al., 2022; Xie et al., 2022)}).
% \end{itemize}

Contexts with multiple citation marks are more complex and require special parsing and handling, especially when citation marks are mixed or placed in different parts of the sentence. A full list of citation mark formats identified in our data and examples is shown in Table \ref{tab:citation_mark_formats}.

\subsection{Cleansing citation contexts for crowdsourcing}\label{lbl:pre-processing-filtering}

By identifying various styles of citation marks, resolving citation marks referring to the cited (original) paper from multiple citation marks, and analyzing sentence structures, we were able to provide clean citation contexts to crowdworkers. We created a filtered dataset containing only citation contexts that uniquely identify the related cited paper (see Appendix\footnote{\url{https://github.com/lamps-lab/CC30k/blob/main/appendix.pdf}}). A regular expression-based approach (see our GitHub repository\footnote{\url{https://github.com/lamps-lab/CC30k/blob/main/notebooks/R001_AWS_Labelling_Dataset_Preprocessing_Mturk.ipynb}}) was applied to obtain citation contexts in which cited papers are unambiguously identified (Table \ref{tab:citation_mark_formats}: Item (e) is problematic and cannot uniquely identify a cited paper because grouped and separate numeric citation marks appear together, making it unclear which cited paper the text refers to). This process reduced the number of citation contexts from 41,244 to 25,829 (Table \ref{tab:final-dataset}).

\renewcommand{\arraystretch}{1.4} % Increase row 
\begin{table*}[ht]
\small
\centering
\caption{Summary of scenarios in which a citation context contains citation marks referring to uniquely identified cited papers.}
\begin{tabular}{p{0.25\textwidth}p{0.35\textwidth}p{0.25\textwidth}r}
% \begin{tabular}{lllc}
\hline
\textbf{Scenario} & \textbf{Description} & \textbf{Example} & \textbf{Count} \\ \hline
\textbf{Contexts with single citation mark} & These citation contexts are straightforward and easily identifiable & (Ya et al., 2022; Dai \& Hang, 2021) or [1] or [1, 2, 3] & 20,830 \\ %\hline
\textbf{Contexts with multiple APA-like citation marks} & These contexts have more than one citation mark in a single context, which could be challenging for workers without additional details & (Ya et al., 2022; Dai \& Hang, 2021) text text (Zhao et al., 2022) & 4,871 \\ %\hline
\textbf{Contexts that contain only the first author’s name} & These contexts only contain the  first author name of the cited paper within the context & Zhao argues that... or According to Zhang... & 128 \\ \hline
\textbf{Total} & & & \textbf{25,829} \\ \hline

\end{tabular}
\label{tab:final-dataset}
\end{table*}

\subsection{Crowdsourcing annotation}
\subsubsection{Task Design}

We employed crowdworkers recruited through Amazon Mechanical Turk to annotate the dataset (25,829 citation contexts) with three ROS categories. Each context was assigned to three workers; each worker was asked to assign a label independently to a citation context. We applied majority voting (2/3 or 3/3 agreement) to determine the final label for a citation context. To progressively monitor the labeling process, we posted the entire job in 26 batches, each containing about 1000 citation contexts. For each labeling task, we recorded the time in seconds spent on the task, the label, and the worker's ID.

\subsubsection{Types of ROS labels and descriptions}
\begin{itemize}[leftmargin=*]
    \item \textbf{Positive:} The context suggests successful reproducibility or replicability, such as the reuse of data, code, or concepts from the cited paper. It may include terms such as \textit{reproduce}, \textit{replicate}, or \textit{repeat the experiments}, or references to the software or processes from the cited paper being used for pre-processing or comparison.
    \item \textbf{Negative:}  The context hints about irreproducibility or irreplicability, such as the unavailability of the cited paper's data or code, or unsuccessful attempts to obtain the results. 
    \item \textbf{Neutral:}  
    The context simply mentions (cites) a paper without providing any hints about reproducibility. These contexts lack any indication of attempts to run the implementation or verify the results.
\end{itemize}

\begin{figure}[h]
  \centering
  \setlength{\fboxsep}{0pt} % Adjust the padding
  \setlength{\fboxrule}{0.5pt} % Adjust the border width
  \fbox{\includegraphics[width=1\linewidth]{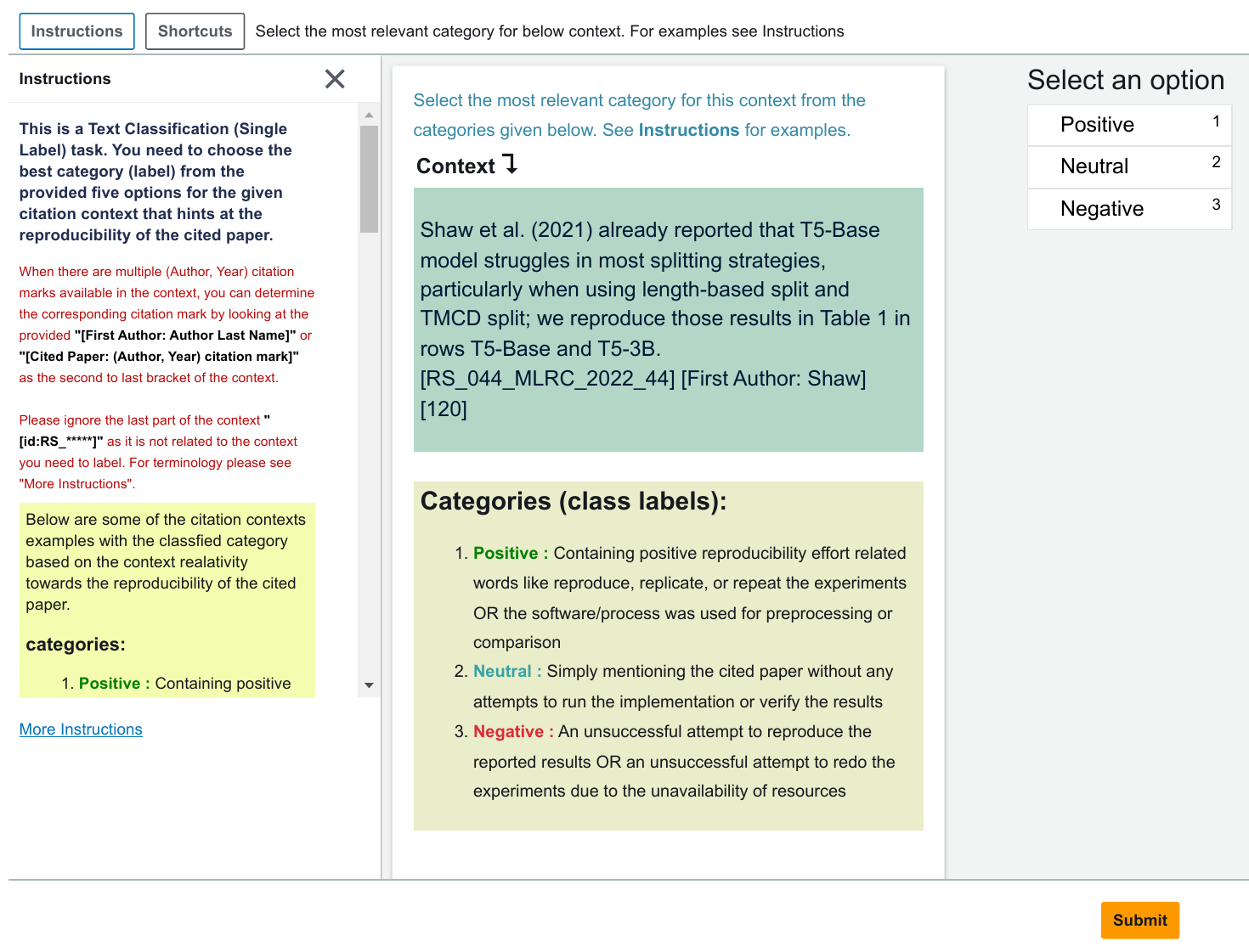}}
  \caption{Crowdsourcing task interface using Crowd HTML elements.}
  \label{fig:labeling_gui}
\end{figure}

\subsubsection{Worker selection mechanism}

To ensure collecting high-quality labels, we implemented an iterative worker selection mechanism using a pilot dataset of 20 citation contexts for which we had ground truth labels, by in-house domain experts. This pilot dataset included 10 actual citation contexts from our finalized dataset and 10 dummy citation contexts, the labels of which could easily be determined.

In the worker selection stage, we posted the pilot dataset as a separate task, requiring at least 10 workers to participate. % in the task. 
To select more reliable workers, we restricted participation to \textit{Mechanical Turk Masters} \cite{mturkSimplifiedMasters}, a program under MTurk for selecting a specialized group of workers. To build a large pool of qualified workers, we repeatedly posted the pilot dataset in multiple batches, allowing only new workers who had not annotated in previous batches to annotate a new batch. Through this iterative process, we built a pool of 138 workers.

We evaluated the workers' responses against our ground truth labels and selected 16 workers who:
\begin{itemize}
    \item labeled at least 15 citation contexts from the pilot dataset;
    \item achieved an accuracy of over 90\% in their answers; and
    \item correctly labeled all the dummy citation contexts.
\end{itemize}
We designed a graphic interface for the labeling task shown in Fig.~\ref{fig:labeling_gui} using the Mechanical Turk Crowd HTML Elements.

%%\subsubsection{Interface Design}

%%We designed a custom template for the labeling task GUI using \textit{Mechanical Turk Crowd HTML Elements} \cite{mturk-crowd-elements}, which abstracts HTML markup, CSS, and JavaScript functionality into HTML tags. %This allowed for an efficient and visually consistent interface. As shown in Figure \ref{fig:labeling_gui}, the GUI included a %section containing detailed instructions for workers, along with examples for each label category. We provided definitions for key scientific terms, e.g., \textit{citation}, \textit{citation context}, and \textit{cited paper}. The GUI iteratively displayed each citation context each with a label selection area on the right. Workers were required to choose a label from three options (\textit{Positive}, \textit{Negative}, \textit{Neutral}) and click the \textit{Submit} button to proceed to the next context. The instructions, examples, and definitions were consistently presented for every task to reinforce understanding and minimize errors. It took approximately 27 days to complete the labeling. 
% We used \textit{VS Code} as the Integrated Development Environment (IDE) to develop the template, and the \textit{Mechanical Turk Developer Sandbox} \cite{mturk-Sandbox-Developer} (a simulated environment for testing applications and tasks prior to publication) to test the GUI's functionality before deployment.

\subsection{Augmenting Negative Citation Contexts}\label{sec:negative-augmentation}

During the crowdsourcing stage, we identified a severe imbalance in the distribution of ROS labels, with negative contexts representing only $\sim$1\% of the dataset (Table~\ref{tab:summary_statistics}),  due to the scarcity of explicit negative citation contexts. To address this issue, we applied an AI-assisted approach to add more negative citation contexts to balance the positive ROS labels. 

Using S2GA, we first collected 692{,}604 citation contexts from 21{,}757 computer science papers published after 2017, consistent with the lower bound year used in our crowdsourced set. We then classified these contexts using an ensemble model consisting of 5 transformer models (\textit{SPECTER \cite{cohan-etal-2020-specter}}, \textit{SciBERT \cite{beltagy-etal-2019-scibert}}, \textit{DistilBERT \cite{sanh2020distilbertdistilledversionbert}}, \textit{BioBERT \cite{10.1093/bioinformatics/btz682}}, and \textit{BlueBERT \cite{peng-etal-2019-transfer}}), each fine-tuned using the crowdsourcing labeled data.

Before classifying the $\sim$600k contexts, we evaluated\footnote{\url{https://github.com/lamps-lab/CC30k/blob/main/notebooks/R001_Extend_CC25k_Dataset.ipynb}} all models, including the ensemble, on the test set reported in Table~\ref{tab:mturk_validation}. The ensemble model achieved a weighted average F1 score of 0.81 and produced 43{,}790 negative candidates.

\begin{figure}[h]
  \centering
  \setlength{\fboxsep}{0pt} % Adjust the padding
  \setlength{\fboxrule}{0.5pt} % Adjust the border width
  \fbox{\includegraphics[trim=18 4 16 5, clip, width=0.98\linewidth]{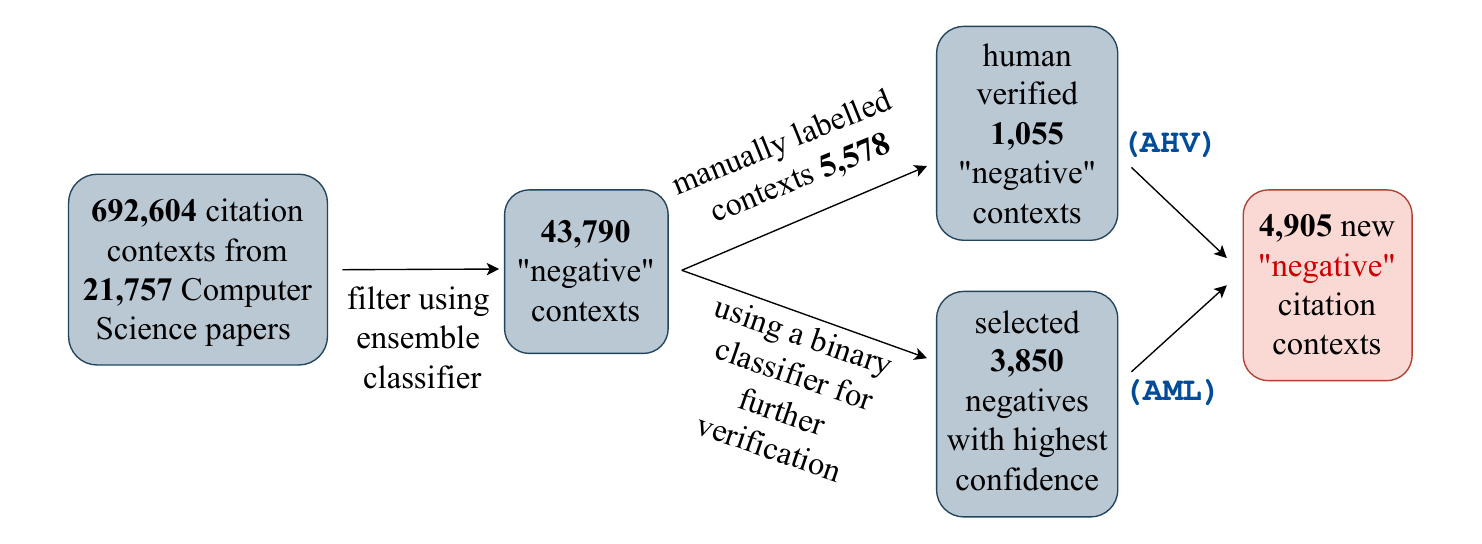}}
  \caption{The pipeline of augmenting negative citation contexts; AML: augmented and machine labeled, AHV: augmented and human-validated.}
  \label{fig:negative_augmenting}
\end{figure}

From these, we manually verified a random sample of 5{,}578 contexts, and ultimately identified 1{,}055 negative citation contexts. However, this is still insufficient to balance the positive labels. From the remaining verified 4,523 non-negative contexts, we randomly selected 1,055 as additional ground truth samples. Using this data (a total of 2,110 samples) we fine-tuned several binary classifiers, including SciBERT, RoBERTa, DistilBERT, DeBERTa, and GPT-4o in zero-shot and few-shot settings, to identify actual negatives. RoBERTa achieved the highest F1 score of 0.67, and we used this model as a verification step to filter negative candidates with high confidence. Specifically, we fine-tuned a RoBERTa \cite{liu2019robertarobustlyoptimizedbert, cheang2020languagerepresentationmodelsfinegrained} binary classifier (negative vs. non-negative) and applied the model to the remaining 38,212 candidate negatives, resulting in 3,850 high-confidence ($>$0.99) candidates. Combined with the manually verified subset, this process yielded 4,905 new negative contexts (Fig.~\ref{fig:negative_augmenting}), which were added to the CC30k dataset.

We also introduced a metadata field, \texttt{label\_type}, which indicates the origin of the label. A label may be obtained through \textit{crowdsourcing} (\texttt{crowdsourced}), through human validation of negatively augmented examples (\texttt{augmented\_human\_validated}), or through automatic assignment via the augmentation pipeline without human validation (\texttt{augmented\_machine\_labeled}).
The statistics of the final dataset is presented in Table \ref{tab:summary_statistics}, and the metadata schema is shown in Table~\ref{tab:dataset_columns}.

\section{Data Records}

The CC30k dataset is in a structured format to facilitate easy use for 
analysis and modeling. The dataset is stored as a single CSV file, with each row representing a unique citation context and its labels and associated metadata of the cited and the citing paper. 
The GitHub repository\footnote{\url{https://github.com/lamps-lab/CC30k/blob/main/README.md}} includes a README file with detailed instructions on how to use the dataset and a description of its structure. The dataset contains a CSV file with 37 columns described in the Table \ref{tab:dataset_columns}. 

\renewcommand{\arraystretch}{1.25}
\begin{table*}[h!]
% \small
\centering
\caption{The fields in the CC30k dataset. Metadata of citing and cited papers are collected from S2GA.}
\begin{tabular}{p{0.26\textwidth}p{0.68\textwidth}}
\hline
\textbf{Column Name} & \textbf{Description} \\ \hline
\texttt{input\_index} & Unique ID for each citation context. \\ %\hline
\texttt{input\_context} & Citation context that workers are asked to label. \\ %\hline
\texttt{input\_file\_key} & Identifier linking the context to a rep-study. \\ %\hline
\texttt{input\_first\_author} & Name or identifier of the first author of the cited paper. \\ %\hline
\texttt{worker\_id\_w1} & Unique ID of the first worker who labeled this citation context. \\ %\hline
\texttt{work\_time\_in\_seconds\_w1} & Time (in seconds) the first worker took to label the citation context. \\ %\hline
\texttt{worker\_id\_w2} & Unique ID of the second worker who labeled this citation context. \\ %\hline
\texttt{work\_time\_in\_seconds\_w2} & Time (in seconds) the second worker took to label the citation context. \\ %\hline
\texttt{worker\_id\_w3} & Unique ID of the third worker who labeled this citation context. \\ %\hline
\texttt{work\_time\_in\_seconds\_w3} & Time (in seconds) the third worker took to label the citation context. \\ %\hline
\texttt{label\_w1} & Label assigned by the first worker. \\ %\hline
\texttt{label\_w2} & Label assigned by the second worker. \\ %\hline
\texttt{label\_w3} & Label assigned by the third worker. \\ %\hline
\texttt{batch} & Batch number for the posted Mechanical Turk job. \\ %\hline
\texttt{majority\_vote} & Final label based on the majority vote among workers’ labels (\textbf{\footnotesize reproducibility-oriented sentiment}). \\ %\hline
\texttt{majority\_agreement} & Indicates how many of the three workers agreed on the final majority vote. \\ %\hline
\texttt{rs\_doi} & Digital Object Identifier (DOI) of the reproducibility study paper. \\ %\hline
\texttt{rs\_title} & Title of the reproducibility study paper. \\ %\hline
\texttt{rs\_authors} & List of authors of the reproducibility study paper. \\ %\hline
\texttt{rs\_year} & Publication year of the reproducibility study paper. \\ %\hline
\texttt{rs\_venue} & Venue (conference or journal) where the reproducibility study was published. \\ %\hline
\texttt{rs\_selected\_claims} & Number of claims selected from the original paper for reproducibility study (by manual inspection). \\ %\hline
\texttt{rs\_reproduced\_claims} & Number of selected claims that were successfully reproduced (by manual inspection). \\ %\hline
\texttt{reproducibility\_label} & Reproducibility label assigned to the original paper based on the number of \texttt{rs\_reproduced\_claims} \textit{(reproducible, not-reproducible, partially-reproducible [if 0 $<$ \textbf{rs\_reproduced\_claims} $<$ rs\_selected\_claims] )}. \\ %\hline
\texttt{org\_doi} & DOI of the original (cited) paper that was assessed for reproducibility. \\ %\hline
\texttt{org\_title} & Title of the original (cited) paper. \\ %\hline
\texttt{org\_authors} & List of authors of the original (cited) paper. \\ %\hline
\texttt{org\_year} & Publication year of the original (cited) paper. \\ %\hline
\texttt{org\_venue} & Venue where the original (cited) paper was published. \\ %\hline
\texttt{org\_paper\_url} & URL to access the original (cited) paper. \\ %\hline
\texttt{org\_citations} & Number of citations received by the original (cited) paper (collected by the date 2024/02/15). \\ %\hline
\texttt{citing\_doi} & DOI of the citing paper that cited the original (cited) paper. \\ %\hline
\texttt{citing\_year} & Publication year of the citing paper. \\ %\hline
\texttt{citing\_venue} & Venue where the citing paper was published. \\ %\hline
\texttt{citing\_title} & Title of the citing paper. \\ %\hline
\texttt{citing\_authors} & List of authors of the citing paper. \\
\texttt{label\_type} & label source: \texttt{crowdsourced} or \texttt{augmented\_human\_validated} or \texttt{augmented\_machine\_labeled}
\\ \hline
\end{tabular}
\label{tab:dataset_columns}
\end{table*}

\begin{figure}[h]
  \centering
  \setlength{\fboxsep}{0pt} % Adjust the padding
  \setlength{\fboxrule}{0.0pt} % Adjust the border width
  % \fbox{\includegraphics[trim=0 20 0 20, clip, width=0.6\linewidth]{figures/word_count_distribution_.pdf}}
  \fbox{\includegraphics[width=0.82\linewidth]{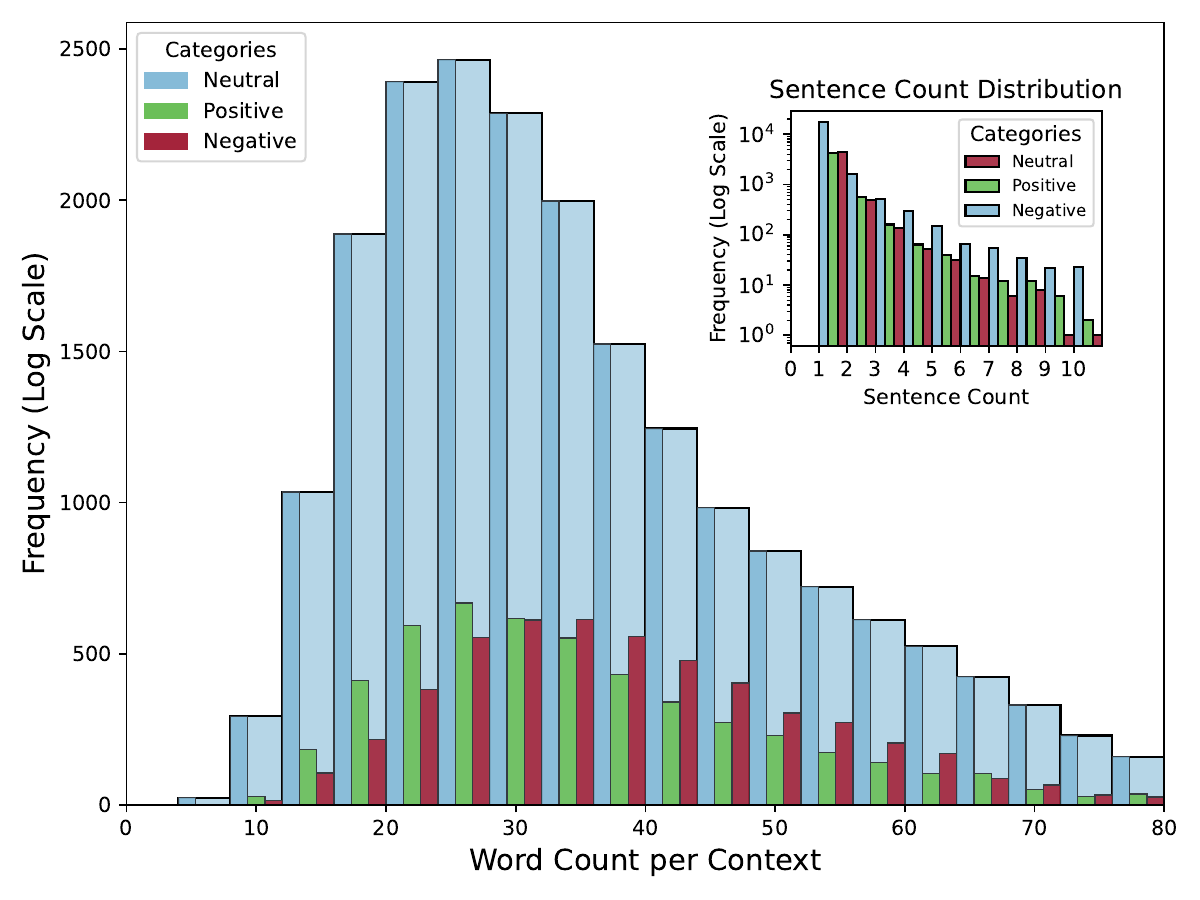}}
  \caption{Word count per context distribution for different categories: \textit{Neutral}, \textit{Positive}, and \textit{Negative} (based on the entire CC30k dataset).}
  \label{fig:word_count_distribution}
\end{figure}

We report the average word count and the average number of sentences per context, for each sentiment category (Table \ref{tab:summary_statistics}). Fig. \ref{fig:word_count_distribution} shows these distributions.

%\subsection{Distribution of Citation Contexts, Citing Papers, and ROS Proportions}
\subsection{Labeling Outcomes}
\begin{figure}[h]
  \centering
  \setlength{\fboxsep}{0pt} % Adjust the padding
  \setlength{\fboxrule}{0.0pt} % Adjust the border width
  % \fbox{\includegraphics[trim=0 20 0 20, clip, width=0.6\linewidth]{figures/word_count_distribution_.pdf}}
  \fbox{\includegraphics[width=0.8\linewidth]{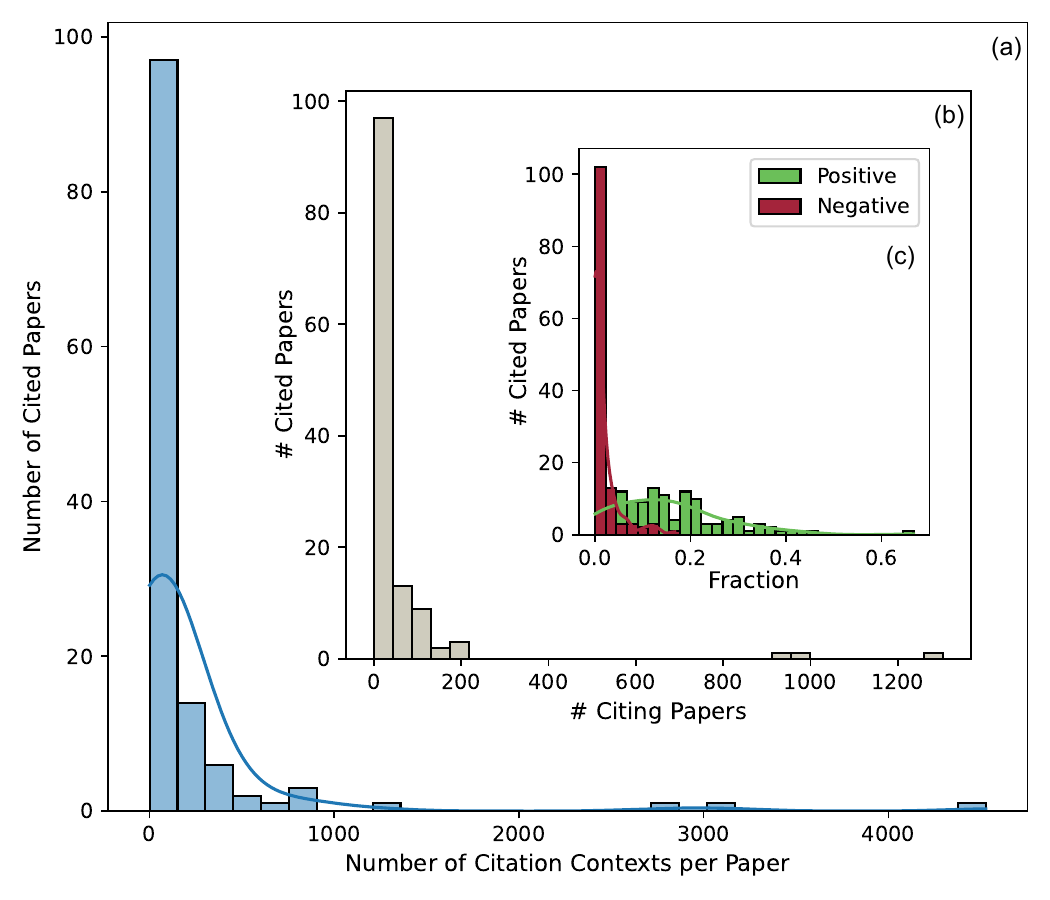}}
  \caption{Distribution of citation contexts, citing papers, and sentiment proportions across cited papers (crowdsourced portion): (a) Citation context count distribution across cited papers, (b) Citing paper count distribution across cited papers, (c) Distribution of the fraction of positive and negative citation contexts among all citation contexts in a paper.}
  \label{fig:citation_contexts_distribution}
\end{figure}

Fig. \ref{fig:citation_contexts_distribution} shows the distribution of citation contexts and citing papers across cited papers, along with the fraction of positive and negative citation contexts for crowdsourced portion. The subfigures in Fig.~\ref{fig:citation_contexts_distribution} reveal a highly skewed distribution across the three dimensions. In Fig. \ref{fig:citation_contexts_distribution}(a), the majority of cited papers are cited in no more than 100 citation contexts, while a few (2.36\%) are cited in more than 2000 contexts, which are likely to be works that impact a large number of papers. A similar trend is evident in Fig.~\ref{fig:citation_contexts_distribution}(b). Fig. \ref{fig:citation_contexts_distribution}(c) shows that the majority of cited papers receive less than 5\% negative ROS citations, whereas positive citation contexts exhibit a wider distribution.

\begin{table}[h]
\footnotesize
\centering
\caption{Summary Statistics of CC30k Dataset after Negative Augmentation (30,734 Contexts) C:Crowdsourced, AHV: Augmented and Human-Validated, AML: Augmented and Machine Labeled}
\label{tab:summary_statistics}
\begin{tabular}{lrrcc}
\hline
\textbf{\footnotesize Category} & \textbf{\footnotesize \# Contexts} & \textbf{\%} & \textbf{\footnotesize Avg. Words} & \textbf{\footnotesize Avg. Sent.} \\ \hline
\textit{Positive (C)} & 5,102 & 16.63\% & 36.55 & 1.33 \\ 
\textit{Neutral (C)} & 20,448 & 66.65\% & 36.03 & 1.29 \\ 
\textit{Negative (C)} & 279 & 0.91\% & 35.84 & 1.20 \\ \hline
\textit{Negative (AHV)} & 1,055 & 3.26\% & 38.89 & 1.22 \\ 
\textit{Negative (AML)} & 3,850 & 12.55\% & 39.06 & 1.26 \\ \hline
\textbf{Total} & 30,734 & 100\% & 36.59 & 1.29 \\  
\hline
\end{tabular}
\end{table}

% -------------------------
% statistical summary

\vspace{-7pt}

\renewcommand{\arraystretch}{1.25}
\begin{table}[ht]
\small
    \centering
    \caption{Macro average performance metrics of five popular open-source sentiment classification models available on HuggingFace (3-class) compared to ground truth from MTurk labels (support: Negative = 279, Neutral = 20,448, Positive = 5,102; total = 25,829).}
    \label{tab:classification-reports}
    \begin{tabular}{lccc}
        \hline
        \textbf{Model} & \textbf{Precision} & \textbf{Recall} & \textbf{F1-Score} \\
        \hline
        RoBERTa-based model \cite{hartmann2021} & 0.35 & 0.44 & 0.37 \\
        BERT-based model \cite{Seethal-Hugging} & 0.36 & 0.37 & 0.36 \\
        BERTweet \cite{finiteautomata-Hugging} & 0.38 & 0.48 & 0.40 \\
        BERT AutoTrain \cite{Souvikcmsa-Hugging} & 0.36 & 0.50 & 0.37 \\
        BERT sbcBI \cite{SbcBI-Hugging} & 0.34 & 0.53 & 0.30 \\
        \hline
    \end{tabular}
\end{table}

\section{Data Quality Assessment} \label{technical_validation}

Table~\ref{tab:summary_statistics} shows that the final dataset contains crowdsourced data, augmented and human-validated (AHV), and augmented and machine-labeled (AML). The AHV data are human-validated and used as a gold standard to train/evaluate supervised models. The AML portion only contains citation contexts with negative ROS labels. Here, we focus on assessing the quality of the crowdsourced data by manually validating the labels using a subset containing 1\% of samples selected from the crowdsourced portion. To avoid selection bias due to the imbalanced distribution across three ROS labels, we adopted a stratified sampling approach, in which we selected 10 majority-voted citation contexts from each batch of 1,000 citation contexts with 4 positive, 3 negative, and 3 neutral labels.  Specifically, we randomly sampled 3 Neutral contexts per batch (2 with a majority voting agreement of 2/3 and 1 with an agreement of 3/3), 4 Positive contexts per batch (2 with agreement of 2/3 and 2 with agreement of 3/3), and 3 Negative contexts per batch (2 with agreement of 2/3 and 1 with agreement of 3/3). This results in a subset of 244 citation contexts.  
Table \ref{tab:mturk_validation} provides a summary of the comparison between the manually-validated and the majority-voted labels.

\renewcommand{\arraystretch}{1.25}
\begin{table}[h]
\small
\centering
\caption{Comparison of ground truth labels with Mechanical Turk annotations for the selected sample}
\label{tab:mturk_validation}
\begin{tabular}{lcc}
\hline
\textbf{Label Category} & \textbf{Manually-validated} & \textbf{Majority-voted} \\ \hline
Negative       & 55                      & 55                    \\ %\hline
Positive       & 102                     & 99                    \\ %\hline
Neutral        & 87                      & 75                    \\ \hline%\Xhline{3\arrayrulewidth}
Total          & 244                     & 229                   \\ \hline
\end{tabular}
\end{table}
The discrepancy between the two total numbers is due to 15 mislabels, as determined by majority voting, resulting in an overall accuracy of 93.85\% for the crowdsourced data. All the mislabeling was limited to scenarios with the majority voting agreement of 2/3. No mislabels were found in cases where all three workers agreed on a label (3/3 agreement). 
To further assess the quality of the crowdsourced labels, we calculated the precision, recall, and F1-score for each label category (see Table \ref{tab:evaluation_metrics}). The macro- and weighted-average F1-scores are both 94\%, indicating a high quality of the crowdsource data. \rr{%Overall, annotators showed a high percent agreement (99.35\%). Krippendorff's $\alpha$ (ordinal) was 0.29 due to the extreme skew in label distribution, interpreted as ``fair'' on the Landis \& Koch \cite{Landis1977-ou} scale. %The lower $\alpha$ reflects the extreme skew in label distribution, as most items were labeled Neutral. 
The overall percent agreement among annotators was 99.35\%. We also computed Krippendorff's $\alpha$ (ordinal) to quantify inter-annotator reliability. According to the Landis \& Koch \cite{Landis1977-ou} scale, $\alpha = 0.29$ would be interpreted as ``fair'' agreement. However, this lower $\alpha$ value is primarily due to the extreme skew in label distribution, with the majority of items labeled as Neutral.
}

\begin{table}[h]
\small
\centering
\caption{Evaluation metrics for annotation quality}
\label{tab:evaluation_metrics}
\begin{tabular}{lcccc}
% \begin{tabular}{|l|c|c|c|c|}
\hline
\textbf{Label Category} & \textbf{Precision} & \textbf{Recall} & \textbf{F1-Score} & \textbf{Support} \\ \hline
Negative                & 0.89             & 1.00             & 0.94            & 55             \\ %\hline
Neutral                 & 0.96             & 0.86          & 0.91            & 87             \\ %\hline
Positive                & 0.95             & 0.97          & 0.96            & 102            \\ \hline %\Xhline{3\arrayrulewidth}
% \textbf{Accuracy}       & \textbf{0.9385}    & \textbf{0.9385} & \textbf{0.9385}   & \textbf{244.0}   \\ \hline
\textbf{Macro Avg.}               & \textbf{0.93}             & \textbf{0.94}          & \textbf{0.94}            & \textbf{244}            \\ 
\textbf{Weighted Avg.}           & \textbf{0.94}             & \textbf{0.94}          & \textbf{0.94}            & \textbf{244}            \\ \hline
% Micro-Avg               & 0.94             & 0.94          & 0.94            & 244            \\ \hline
\end{tabular}
\end{table}

% \vspace{-8pt}
\section{Potential Uses of the Dataset}
\subsection{Baseline performance of off-the-shelf sentiment classifiers}

Using the labeled citation contexts, we first evaluate the performance of five open-source sentiment classification models on  \textit{HuggingFace} \cite{huggingfaceModelsHugging}. Performance measures in Table~\ref{tab:classification-reports}, reveal that these models perform poorly when evaluated on the crowdsourced portion of citation contexts in our dataset. The macro-averaged F1-scores for all models are below 0.41. The results highlight the limitations of off-the-shelf sentiment models in addressing complex and domain-specific tasks. Our CC30k dataset fills this gap by providing a specialized resource tailored to ROS predictions. 

\subsection{Fine-tuning large language models}\label{sec:llm-utility}

To showcase the practical utility of CC30k for downstream NLP applications, we conducted a series of experiments in which we fine-tuned base LLMs on our dataset for the task of ROS classification. All models are evaluated using the same test set containing 244 samples with ground truth labels described in section \ref{technical_validation}. We selected two open-source LLMs, LLaMA-3-8B\footnote{\url{https://huggingface.co/NousResearch/Meta-Llama-3-8B-Instruct}} and Qwen-1.5-7B\footnote{\url{https://huggingface.co/Qwen/Qwen1.5-7B}}, for fine-tuning, and one commercial LLM, GPT-4o, for Retrieval-Augmented Generation (RAG). We evaluated eight scenarios for each LLM, including two direct inference (base model) in zero-shot and few-shot (n=5 i.e., five randomly selected examples per category, 15 total) prompting methods, and six fine-tuned models with three training set sizes (3k, 9k, 15k) under both zero-shot and few-shot prompting methods. For example, 3k means we selected 3,000 citation contexts with labels from CC30k in total, with 1,000 from each ROS category. 

\begin{figure*}[h]
  \centering
  \setlength{\fboxsep}{0pt} % Adjust the padding
  \setlength{\fboxrule}{0.0pt} % Adjust the border width
  \fbox{\includegraphics[trim=0 10 0 2, clip, width=1\linewidth]{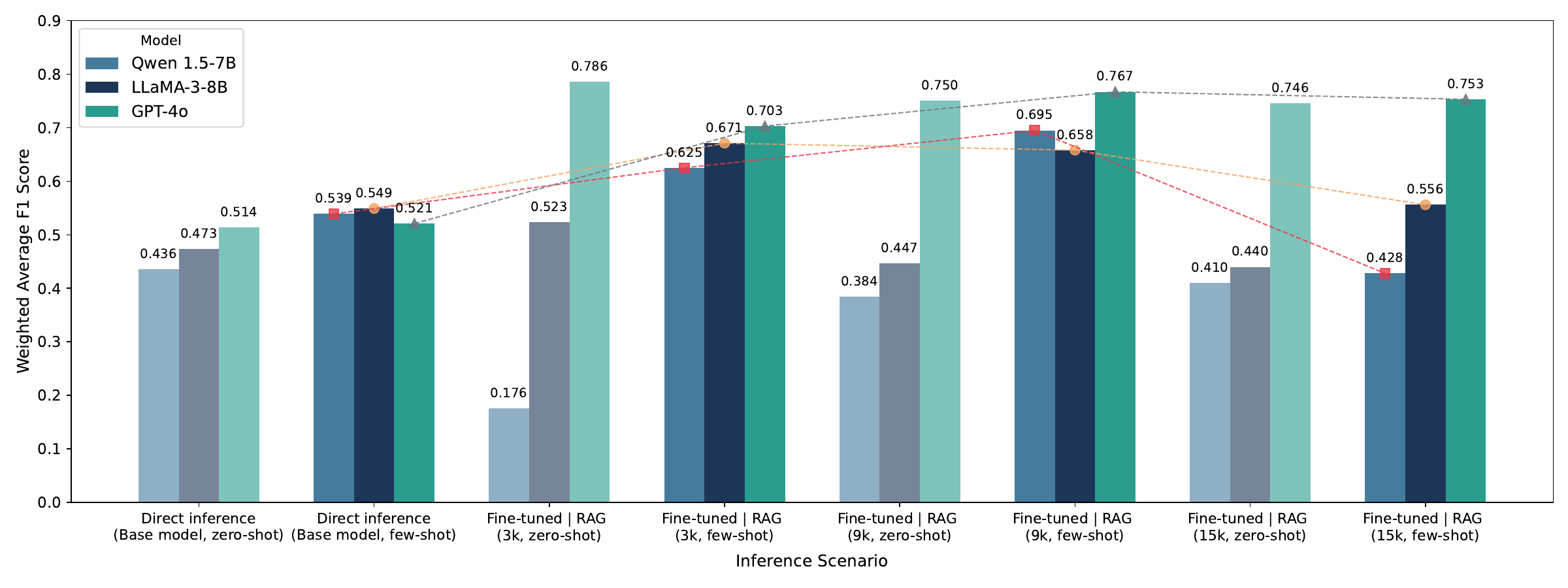}}
  \caption{Reproducibility-Oriented Sentiment Classification Performance of Fine-Tuned LLMs with Varying Training Sample Sizes from the CC30k Dataset. ``3k" indicates 3,000 citation contexts selected from CC30k, with 1,000 from each ROS category. Dashed lines connect few-shot inference scenarios to show model performance trends across increasing fine-tuning data sizes (e.g., the red dashed line tracks Qwen performance in the few-shot settings).}
  \label{fig:llm_finetuning_results}
\end{figure*}

Across models we observe that fine-tuning on CC30k yields consistent and substantial gains over base direct inference in most settings (Fig. \ref{fig:llm_finetuning_results}). The magnitude of the gains depends on the base model, the prompting method, and the fine-tuning data size. Here, we summarize performance gains in the form of weighted average F1-scores for each LLM.

\begin{itemize}
  \item \textbf{Qwen1.5–7B:} In the base setting, the zero-shot score was 0.436, while the few-shot score reached 0.539. Fine-tuning with few-shot prompting improved performance to 0.625 with 3k training samples and 0.695 with 9k training samples, but the score dropped to 0.428 with 15k training samples. The best observed F1-score was \textbf{0.695}, achieved with 9k training samples under few-shot prompting.
  \item \textbf{LLaMA 3–8B:} Fine-tuning with few-shot prompting achieved 0.671 with 3k training samples, which is the highest F1-score, followed by 0.658 with 9k training samples, with lower scores observed with 15k training samples. 
  \item \textbf{GPT-4o (with RAG):} The F1 score peaked at \textbf{0.786} with 3k training samples under zero-shot prompting. The second best F1-score is 0.767 with 9k training samples under few-shot prompting. 
\end{itemize}

% Comparing base vs.\ fine-tuned models.
In summary, fine-tuning on CC30k generally improves performance relative to direct base model inference by 5\% to 27\%. For Qwen and LLaMA, the most evident improvements appear when fine-tuning is combined with few-shot prompting (e.g., Qwen: 0.539 $\rightarrow$ 0.695; LLaMA: 0.549 $\rightarrow$ 0.671). . This demonstrates that parameter updates informed by CC30k plus a small number of in-context examples can be complementary. For GPT-4o, however, RAG fine-tuning produced very large gains even with zero-shot inference (0.514 $\rightarrow$ 0.786 at 3k RAG). 
These observations indicate that CC30k is effective for improving the LLM's performance on ROS classification. However, the best setting is model-dependent, and our dataset provides a sufficient number of samples for fine-tuning. 

% Comparing zero-shot vs.\ few-shot prompting.
In many cases, few-shot prompting improves performance over zero-shot, consistent with prior findings that in-context examples help guide model outputs by conditioning model behavior on task-specific demonstrations \cite{NEURIPS2020_1457c0d6, lampinen-etal-2022-language}. 

However, when the model is equipped with strong external context via RAG, zero-shot inference can match or even exceed few-shot performance (the GPT-4o 3k RAG zero-shot case is an example: 0.786 zero-shot vs.\ 0.703 few-shot). An explanation is that retrieval surfaces precise supporting passages that eliminate the need for additional in-prompt examples. These phenomena are aligned with several empirical and theoretical analyses of in-context learning and with the benefits of retrieval-augmented approaches \cite{liu-etal-2022-makes, gao2024retrievalaugmentedgenerationlargelanguage}.

We note that the performance improvements are not strictly monotonic with more fine-tuning data for every model. For example, Qwen's few-shot F1-score increases from 3k to 9k but then drops at 15k; LLaMA shows a similar pattern. One plausible contributor is label noise in our training contexts: crowdsourced annotations contain noise due to the ambiguity in the text, subtle inter-annotator differences, and overlapping categories, and label noise is known to cause unstable fine-tuning patterns and occasional degradation as training sample sizes grow if noise is not explicitly addressed. 

Thus, the non-monotonic behavior observed can be explained primarily by the model sensitivity to noisy supervision during parameter updates. For GPT-4o, which we could only use in a retrieval-augmented generation (RAG) setup due to the model weights not being publicly available, additional interactions between retrieval and in-context signals with imperfect labels may also contribute. Survey and empirical work \cite{Song2023-ps, Liang_2022} on learning with label noise supports this interpretation.

\subsection{Other Usages}  
In addition to improving the performance of LLMs via fine-tuning, the citation contexts in the CC30k dataset can be further linked to citing and cited papers, putting ROS-labeled citation contexts in the citation graph. CC30k can then be used to study patterns in citation practices in AI and potentially other disciplines. Such analyses can reveal disciplinary differences in reproducibility practices and concerns, and provide insights into how reproducibility is discussed and referenced in scholarly publications. Because certain citation contexts may contain explanations of why a cited paper is not reproducible/replicable, CC30k can support identifying key factors influencing AI reproducibility/replicability. This can provide researchers and policymakers with actionable insights into improving reproducibility in science. In particular, because certain citation contexts mention data and software availability/accessibility, CC30k can then be used for reproducibility-aware data and/or software recommendation \cite{he2010contextawarecitation}.

\vspace{-8pt}
\section{Conclusion}
In conclusion, the CC30k dataset is built to address the challenges of reproducibility-oriented sentiment analysis within scientific literature in AI. By providing a carefully labeled, reproducibility-oriented, high-quality dataset of 30,734 citation contexts, this work fills a critical gap in resources for studying reproducibility and replicability in AI research. The creation of the dataset, supported by a robust crowdsourcing pipeline and thorough validation, ensures its reliability and utility for potentially impactful applications. Our evaluation of existing sentiment analysis models highlights the limitations of generic approaches in this domain, underscoring the need for specialized datasets such as CC30k. By making the dataset publicly available, we encourage further exploration and refinement of ROS analysis, promoting trustworthiness in scholarly communications.

%Bibliography
\bibliographystyle{unsrt}  
\bibliography{references}  

\clearpage

\end{document}